\documentstyle[11pt,epsfig]{article}
\font\tenrm=cmr10

 1
\font\elevenrm=cmr10 scaled\magstep 1

\renewenvironment{thebibliography}[1]
 { \elevenrm
   \begin{list}{\arabic{enumi}.}
    {\usecounter{enumi}     \setlength{\parsep}{0pt}
     \setlength{\itemsep}{3pt} \settowidth{\labelwidth}{#1.}
     \sloppy
    }}{\end{list}}

\begin{document}
\title{\Large \bf Mathematical aspects of the nuclear glory phenomenon:\\ from 
backward focusing to Chebyshev polynomials}

\author{V.B.~Kopeliovich$^{a,b}$\footnote{{\bf e-mail}: kopelio@inr.ru}\\
\small{\em a) Institute for Nuclear Research of RAS, Moscow 117312, Russia} \\
\small{\em b) Moscow Institute of Physics and Technology (MIPT), Dolgoprudny, 
Moscow district, Russia} }
\maketitle
{\rightskip=2pc
 \leftskip=2pc
 \noindent}
{\rightskip=2pc
 \leftskip=2pc

\tenrm\baselineskip=11pt
\begin{abstract}
{The angular dependence of the cumulative particles production off nuclei near the kinematical boundary
for multistep process is defined by characteristic polynomials in angular variables $J_N^2(z_N^\theta)$,
where $\theta$ is the polar angle defining the momentum of the final (cumulative) particle, $z_N^\theta = cos (\theta/N)$,  
the integer $N$ being the multiplicity of the process (the number of interactions).
Physical argumentation, exploring the small phase space method, leads to the appearance of equations for these polynomials
$J_N^2[cos(\pi/N)]=0$. The recurrent relations between polynomials with different $N$ 
are obtained, and their connection with known in mathematics Chebyshev polynomials of 2-d kind is established. 
As a result of this equality, differential cross section of the cumulative particle production has characteristic behaviour 
$d\sigma \sim 1/ \sqrt {\pi - \theta}$ at $\theta \sim \pi$ (the backward focusing effect).
Such behaviour takes place for any multiplicity of 
the interaction, beginning with $n=3$, elastic or inelastic (with resonances excitations in intermediate states), and can be called 
the nuclear glory phenomenon, or 'Buddha's light' of cumulative particles.} 
\end{abstract}
 \noindent
\vglue 0.1cm}
\section{Introduction} 
The particles production processes in high energy interactions
of different projectiles with nuclei, in regions forbidden by kinematics for the 
interaction with a single free nucleon (cumulative production processes), are studied intensively since 70-th 
mostly in Dubna (JINR), beginning with the paper \cite{baldin1}, and in Moscow (ITEP) \cite{leksin1,fran-le-2},
some restricted review of data can be found in \cite{koma1,koma2}.

The interpretation of these phenomena as being manifestation of internal structure of
nuclei assumes that the secondary interactions, or, more generally, multiple
interactions processes (MIP) do not play a crucial role in such production. 
In the case of the large angle particle production the background processes which mask 
the possible manifestations of nontrivial features of nuclear structure, are 
subsequent multiple interactions with nucleons inside the nucleus leading to the
particles emission in the "kinematically forbidden" region (KFR).
 In the case of cumulative particles production studies of
multistep, or cascade processe have been unpopular among physisists, because the main
goal of experiments in this field was to reveal manifestations of new nontrivial
effects in nuclear structure. However, it has been proved  \cite{konkop,kop2,bv-res-1,long}
that multistep processes provide certain, not negligible contribution to the cumulative production
cross sections, although other important contributions are not excluded and remain to be the
main purpose of futher studies.

The small phase space method developed previously in \cite{kop2,long} allows to get analytical expressions for the
probability of the multiple interaction processes near the corresponding kinematical boundaries.
The quadratic form in angular variables (deviations from the optimal kinematics) plays the key role
in this approach. The recurrent relations for the characteristic polynomials in polar angles deviations
have been obtained in \cite{koma2} and are reproduced in present talk. 
The connection of these polynomials with Chebyshev polynomials of 2-d kind,
known in mathematics since middle of 19-th century \cite{cheb-engl,cheb-r} and used in approximation 
theory, has been established.
It is an example of interest when physics arguments have led to some results in mathematics. 

In the next section the peculiarities of kinematics of the processes in KFR will be recalled,
in section 3 the small phase space method of the MIP contributions calculation to the particles
production cross section in KFR is described according to \cite{kop2,long}. 
In section 4 the characteristic polynomials in polar angle deviations from the optimal kinematics
are obtained, which define the angular dependence of the cross section on the emission angle of the final (cumulative) 
particle and lead to the backward focusing effect, similar to the known in optics glory
phenomenon. 
The connection of characteristic polynomials with Chebyshev polynomials of 2-d kind
(Chebyshev-Korkin-Zolotarev, or CKZ-polynomials) is established (section 6).
Some generalizations for the case of inelastic rescatterings are discussed as well.
Final section contains conclusions and discussion of problems.

\section{Features of kinematics of the processes in KFR.}  
At large enough incident energy, $\omega_0 = k_0^0 \gg M_f$, we obtain easily
$$ \omega_f - z k_f \leq m_t, \eqno (2.1) $$
which is the basic restriction for such processes. $z=cos\,\theta < 0$ for particle produced
in backward hemisphere, The quantity $(\omega_f - z k_f)/m$, $m$ is the nucleon mass, has been called 
the cumulative number (more precize, the integer part of this ratio plus 1).

For light particles (photon, also $\pi$-meson) iteration of the Compton formula
allows to get the  final energy in the form                     
$$  {1\over \omega_N}- {1\over \omega_0} = {1\over m} \sum_{n=1}^N  \left[1-cos (\theta_n)\right] \eqno (2.2)  $$

The maximal energy of final particle is reached for the coplanar process when
all scattering processes take place in the same plane and each angle equals to
$\theta_k=\theta/N$.
As a result, we obtain after expansion of $cos (\theta/N)$ 
and for large enough $\omega_0$ 
$$\omega_N^{max} \simeq N {2m\over \theta^2} + {m \over 6N}, \eqno (2.3)$$
which works satisfactory beginning with $N=2$.

In the case of the nucleon-nucleon scattering (scattering of particles with equal 
nonzero masses in general case) similar expression can be obtained in somewhat
different way \cite{kop2,long}
$$p_N^{max} \simeq N {2m\over \theta^2} - {m\over 3N}, \eqno (2.4)$$
which works satisfactory for $N>3$ and coincides with previous result for the rescattering of light particles
at large $N$. 

The normal Fermi motion of nucleons inside the nucleus makes these boundaries wider \cite{long}:
$$p_N^{max} \simeq N {2m\over \theta^2} \left[1 + {p_F^{max}\over 2m}\left(\theta + {1\over \theta} \right) \right], \eqno (2.5)$$
at $\theta \sim \pi$, where it is supposed that the final angle $\theta$ is large. For numerical estimates we
took the step function for the distribution in the Fermi momenta of nucleons inside
of nuclei, with $p_F^{max}/m\simeq 0.27$, see \cite{long} and references there.

The elastic rescatterings themselves are only the "top of the iceberg".
Excitations of the rescattered particles, i.e. production of resonances in intermediate
states which go over again into detected particles in subsequent interactions,
provide the dominant contribution to the production cross section \cite{bv-res-1,kop2}.
Simplest examples of such processes may be $NN \to NN^* \to NN$, $\pi N\to \rho N \to \pi N$,
etc.
The relative increase of the final momentum $k_f$ due to resonances excitation - deexcitation
in intermediate states equals approximately
$$ {\Delta k_f\over k_f} \simeq {1\over N} \sum_{l=1}^{N-1} {\Delta M_l^2\over k_l^2}, \eqno (2.6) $$
where $\Delta M_l^2$ is the difference of the masses squared, of the resonance and incident particle.
The point is that the number of diferent processes of this kind grows rapidly with increasing $N$,
like $(N_r +1)^{N-1}$, $N_r$ being the number of resonances.

\section{The small phase space method for the MIP probability calculations.}
This method, most adequate for anlytical and semi-analytical calculations of
the MIP probabilities, has been proposed in \cite{kop2} and developed later in 
\cite{long}. It is based on the fact that, according to established in \cite{kop2}
and presented in previous section kinematical relations, there is a preferable plane of the 
whole MIP leading to the production of energetic particle at large angle $\theta$, 
but not strictly backwards. Also, the angles of subsequent rescatterings are close
to $\theta / N$. Such kinematics has been called optimal, or basic kinematics.
The deviations of real angles from the optimal values are small, they are defined mostly
by the difference $k_N^{max} - k $, where $k_N^{max}(\theta)$ is the maximal possible momentum reachable
for definite MIP, and $k$ is the final momentum of the detected particle.
$k_N^{max}(\theta)$ should be calculated taking into account normal Fermi
motion of nucleons inside the nucleus, and also resonances excitation ---
deexcitation in the intermediate state. Some high power of the difference  $(k_N^{max} - k)/k_N^{max} $
enters the resulting probability.

Within the quasiclassical treatment adequate for our case, the probability product approximation is
valid, and the following starting expression for the inclusive cross section of the particle
production at large angles takes place (see, e.g., Eq. (4.11) of \cite{long}):

$${d^3k\over \omega} f_N = \pi R_A^2 G_N(R_A,\theta) \int {f_1(\vec k_1) d^3k_1\over \sigma^{leav}_1 \omega_1} \prod_{l=2}^N \frac{M_k^2(s_k,t_k) 
\delta(m +\omega_{l-1}-\omega_l-\omega_{l-1})}{(8\pi)^2 \sigma^{leav}_l m k_{l-1} \omega_l\omega_{l-1}} d^3k_l \eqno(3.1) $$
Here $\sigma_l^{leav}$ is the cross section defining the removal (or leaving) of the rescattered object at the corresponding section
of the trajectory. It includes all inelastic cross section, the part of elastic cross section and the part of the resonance
production cross sections, and can be considerably smaller than the total interaction cross section of the $l$-th intermediate
particle with nucleon. $G_N(R_A,\theta)$ is the geometrical factor defining the probability of the $N$-fold interaction with
definite trajectory of the interacting particles (resonances) inside the nucleus. This trajectory is defined mostly
by the final values of $\vec k, \,\theta $, according to the kinematical relations of previous section. $f_1 = \omega_1 d^3\sigma_1/d^3k_1$,
$\omega'_N=\omega$ --- the energy of the observed particle.

After some evaluation and introducing the differential cross sections of binary reactions $d\sigma_l/dt_l(s_l,t_l) $ instead of 
the matrix elements of binary reactions $M_l^2(s_l,t_l)$, we came to the formula for the production cross section 
due to the $N$-fold MIP \cite{kop2,long}
$$f_N(\vec p_0,\vec k)= \pi R_A^2 G_N(R_A,\theta) \int \frac{f_1(\vec p_0,\vec k_1) (k_1^0)^3 x_1^2dx_1 d\Omega_1}{\sigma_1^{leav}\omega_1}
\prod_{l=2}^N\left({d\sigma_l(s_l,t_l)\over dt}\right) \frac{(s_l-m^2-\mu_l^2)^2-4m^2\mu_l^2}{4\pi m\sigma_l^{leav} k_{l-1}} $$
$$\times \prod_{l=2}^{N-1}\frac{k_l^2 d\Omega_l}{k_l(m+\omega_{l-1}-z_l\omega_lk_{l-1})\;}{1\over \omega_N'}\delta(m+\omega_{N-1}-\omega_N-\omega_N').
\eqno(3.2) $$ 

Further details depend on the particular process. For the case of the light particle rescattering, $\pi$-meson for example, $\mu_l^2/m^2\ll 1$,
we have
$${1\over \omega_N'}\delta(m+\omega_{N-1}-\omega_N-\omega_N') = {1\over k k_{N-1}}
\delta\left[ {m\over k} - \sum_{l=2}^N (1-z_l) - {1\over x_1}\left({m\over p_0}+1-z_1\right)\right] \eqno(3.3) $$
To obtain this relation, one should use the equality $ \omega_N' = \sqrt{m^2 + (\vec k_{N-1}^2 - \vec k)^2 } $
(energy-momentum conservation in the last interaction act)
and the known rules for manipulations with the $\delta$-function.
When the final angle $\theta$ is considerably
different from $\pi$, there is a preferable plane near which the whole
multiple interaction process takes place, and only processes near this plane
contribute to the final output (small phase space). 
A necessary step is to introduce azimuthal  deviations from this optimal
kinematics, $\varphi_k$, $k=1,\,...,N-1$.  $\varphi_N=0$ by definition of the plane of the process, $(\vec p_0, \vec k)$.
Polar deviations from the basic values, $\theta/N$, are denoted as $\vartheta_k$, obviously, 
 $\sum_{k=1}^N\vartheta_k = 0$. The direction of the momentum $\vec k_l$ after $l$-th 
interaction, $\vec n_l$,  is defined by the azimuthal angle $\varphi_l$ and the polar angle 
$\theta_l = (l\theta /N) +\vartheta_1+...+\vartheta_l$. Both azimuthal and polar deviations are restricted in this case
by the (small) difference between the value of final spatial momentum $|\vec k|$ (or energy $\omega$) and kinematical boundary
for the corresponding MIP.
At the angle $\theta =\pi$, strictly backwards,
there is azimuthal symmtry, and the processes from the whole interval of azimuthal 
angle $0< \phi < 2\pi$ provide contribution to the final output (azimuthal focusing, see next section).

Then we obtained \cite{kop2,long} making the expansion in $\varphi_l$,  $\vartheta_l$ and including quadratic terms
in these variables:
$$z_k= (\vec n_k \vec n_{k-1}) \simeq cos (\theta/N) (1-\vartheta_k^2/2) -sin (\theta/N) \vartheta_k 
+ sin (k\theta/N) sin [(k-1)\theta/N] (\varphi_k -\varphi_{k-1})^2/2. \eqno (3.4)$$
In the case of the rescattering of light particles the sum enters the phase space of the process
$$ \sum_{k=1}^N (1- cos\vartheta_k) =  N[1-cos(\theta/N)] + cos(\theta/N)
\sum_{k=1}^N\bigg[-\varphi_k^2\, sin^2(k\theta/N) + $$
$$+{1\over cos(\theta/N)}sin(k\theta/N)sin((k-1)\theta/N)\bigg] -{cos(\theta/N)\over 2} \sum_{k=1}^N \vartheta_k^2 \eqno (3.5)$$ 
To derive this equality we used that $\varphi_N=\varphi_0=0$ --- by definition of the plane of the MIP,
and the mentioned relation $\sum_{k=1}^N\vartheta_k = 0$.
We used also the identity,  valid for $\varphi_N=\varphi_0 =0$:
$${1\over 2}\sum_{k=1}^N \left(\varphi_k^2+\varphi_{k-1}^2 \right) sin(k\theta/N)sin[(k-1)\theta/N] = 
cos(\theta/N)\sum_{k=1}^N\varphi_k^2 sin^2(k\theta/N). \eqno (3.5a)$$
It is possible to present it in the canonical form and to perform integration easily, see
Appendix B and Eq. $(4.23)$ of \cite{long}.
As a result, we have the integral over angular variables of the following form:
$$ I_N(\varphi, \vartheta) = \int \delta\biggl[\Delta^{ext} - z_N\bigg(\sum_{k=1}^{k=N} \varphi_k^2 -\varphi_k\varphi_{k-1}/z_N
+\vartheta_k^2/2\biggr)\biggr] \prod_{l=1}^{N-1} d\varphi_ld\vartheta_l = 
 \frac{\left(\Delta^{ext}\right)^{N-2} (\sqrt 2 \pi)^{N-1}}{J_N(z_N) \sqrt N (N-2)! z_N^{N-1}}, \eqno (3.6) $$
Since the element of a solid angle $d\Omega_l= sin(\theta\,l/N)d\vartheta_l d\varphi_l $, we made here substitution
$sin(\theta\,l/N)\,d\varphi_l \to d\varphi_l$.
$z_N= cos(\theta/N)$,  $\Delta^{ext}\simeq m/k - m/p_0 -N(1 - z_N) +(1-x_1)m/p_0 $  
 defines the distance of the momentum (energy) of the emitted particle $\vec k, \, \omega$ from the kinematical boundary for 
the whole $N$-fold MIP.
$$ J_N^2(z) = Det\, ||a_N||, \eqno (3.7)$$
where the matrix $||a||$ defines the quadratic form $Q_N(z)$ which enters the argument of the $\delta$-function in Eq. $(3.6)$:
$$ Q_N(z,\varphi_k) = a_{kl} \varphi_k \varphi_l =\sum_{k=1}^{k=N} \varphi_k^2 -{\varphi_k\varphi_{k-1}\over z}, \eqno (3.8)$$
for example:
$$Q_2 =\varphi_1^2, \quad  Q_3 = \varphi_1^2 +\varphi_2^2 -{\varphi_1\varphi_2 \over cos(\theta/3)}; \quad 
Q_4 = \varphi_1^2 +\varphi_2^2 +\varphi_3^2-{\varphi_1\varphi_2 \over cos(\theta/4)} - {\varphi_2\varphi_3 \over cos(\theta/4)};\; ...$$

The phase space of the process $(3.3)$ which depends strongly on $\Delta_N^{ext}$ after integration over angular 
variables takes the form
$$\Phi_N^{pions} = \int \prod_{l=1}^{N-1} d\Omega_l {1\over \omega_N'}\delta(m+\omega_{N-1}-\omega_N-\omega_N') =
\frac{(\sqrt 2\pi)^{N-1}(\Delta^{ext})^{N-2}}{kk_{N-1}(N-2)!\sqrt N J_N(z_N)z_N^{N-1}}. \eqno(3.9) $$
 
For nucleons rescattering there are some differences from the case of the light particle, but exactly the same quadratic
form in angular deviations comes into consideration \cite{kop2,long,koma1}.

\section{Quadratic form in angular deviations, characteristic polynomials and their properties.}

The obvious recurrent relation takes place for the quadratic form in azimuthal deviations:
$$Q_{N+1}(z,\varphi_k,\varphi_l)  = Q_N(z,\varphi_k,\varphi_l) +\varphi_N^2 -\varphi_N\,\varphi_{N-1}/z, \eqno (4.1)$$
with $z=cos[\theta /(N+1)]$, has the same value in both sides of this equation, $\varphi_{N+1} =0$ by definition 
of the plane of the process.

Let $t$ be the transformation (matrix) which brings our quadratic form to the canonical form:
$ \tilde t\, a\, t = {\cal I}, $
where ${\cal I}$ is the unit matrix $n\times n$, and $\tilde t_{kl} = t_{lk}$.
Then the equality takes place for the Jacobian of this transformation
$$ (det\,||t||)^{-2} = J_N^2(z) = det\,||a_N||, \qquad (det\,||t||)^{-1} = J_N(z) = \sqrt{det\,||a_N||}. \eqno(4.2)$$

It is convenient to present the quadratic form  in $\phi_k, \phi_l$ which enters the $\delta$ - function in $(3.6)$, as
the sum of squares of certain combinations of deviations $\varphi_k$
$$ Q_{N+1}(\varphi_k,\varphi_l) =J_2^2\left(\varphi_1-{\varphi_2\over 2zJ_2^2}\right)^2 +{J_3^2\over J^2_2}
\left(\varphi_2-{J_2^2\varphi_3\over 2zJ_3^2}\right)^2 +...  + {J_N^2\over J_{N-1}^2}\left(\varphi_{N-1}-{J_{N-1}^2\varphi_{N}\over 2z J_{N}^2}\right)^2 +
{J_{N+1}^2\over J_{N}^2} \varphi_{N}^2 \eqno (4.3)$$
with $J_2^2=1$.
From the recurrent relation $(4.1)$ we can write the equality for the last several terms in quadratic form $(4.3)$,
depending on $\varphi_{N-1}$ and $\varphi_N$:
$$ {J_N^2\over J_{N-1}^2} \varphi_{N-1}^2 + \varphi_N^2 -{\varphi_N \varphi_{N-1}\over z} =
{J_N^2\over J_{N-1}^2} \left(\varphi_{N-1} - {J_{N-1}^2\over J_{N}^2}{\varphi_N \over 2 z}\right)^2 + {J_{N+1}^2\over J_{N}^2}\varphi_N^2. \eqno(4.4)$$
From equality of coefficients before $\varphi_N^2$ in the left and right sides we obtain
$$ 1 = {J_{N-1}^2\over 4z^2 J_{N}^2} + {J_{N+1}^2\over J_{N}^2}. \eqno(4.5) $$
The recurrent relation follows immediately:
$$J_{N+1}^2(z)= J_{N}^2(z)-{1\over 4z^2}J_{N-1}^2(z), \eqno (4.6) $$ 
here $z=cos[\theta/(N+1)]$.
The following general formula for $J_N^2(z_N)$ has been obtained in \cite{long}, Eq. (4.23) \footnote{In the paper \cite{kop2}, Eq (15), this
formula has been presented for $N$ up to $N=5$.}:
$$Det ||a_{kl}|| =J_N^2(z_N) = 1 + \sum_{m=1}^{m < N/2}\left(-{1\over 4z_N^2}\right)^m 
{\prod_{k=1}^{m}(N-m-k)\over m!} =  \sum_{m=0}^{m < N/2}\left(-{1\over 4z_N^2}\right)^m C^m_{N-m-1}, \eqno (4.7)$$
$z_N=cos (\theta/N)$, $Det\, ||a_{kl}||$ is the determinant of the matrix $||a||$, $C_n^m$ is the number of combinations.

As it became clear to us recently, the polynomials $(4.7)$ coincide, up to some factor depending on $z=cos (\theta/N)$
with Chebyshev polynomials of 2-d kind, discovered in the middle of 19-th century \cite{cheb-engl, cheb-r}.
The connection of characteristic polynomials $J^2_N\left[cos(\theta/N)\right]$ with Chebyshev polynomials of 2-d kind
(CKZ-polynomials) will be described in next section.

When the final angle $\theta = \pi$, there is no restriction on the one of azimuthal deviations, at least, 
the plane of the whole MIP can be rotated around common axis of symmetry (direction of the incident momentum $\vec p_0$
which coincides with the direction of final momentum $\vec k$), the qudratic form in azimuthal deviations becomes 
degenerate, i.e. $det ||a_N(\theta = \pi)|| = J_N^2(\theta =\pi) =0$

The condition $J_N(\pi/N) =0$ leads to the equation for $z_N^\pi$ which solution
(one of roots) provides the value of $cos(\pi/N)$ in terms of radicals.
The following expressions for some of these Jacobians take place \cite{kop2,long}
$$J_2^2(z)=1; \qquad J_3^2(z) = 1-{1\over 4z^2}; \qquad J_4^2(z)= 1-{1\over 2z^2}, \eqno (4.8) $$
Obviously enough, $J_3(\pi/3)=0$, 
$J_4(\pi/4)=0$; less trivial examples were given in \cite{long,koma1,koma2}.

For arbitrary $N$,  $J_N^2$ is a polynomial in $1/4z^2$ of the power $|(N-1)/2|$ (integer part of $(N-1)/2$.
Since the solutions of the equations $J_N^2(z)=0$ are not known in general form when the power of the polymial is
greater than $5$, the knowledge of at least one solution, $z=cos(\pi/N)$ can be helpful.

These equations can be obtained using the elementary mathematics methods as well,
however, the general expression for arbitrary $N$ may be of interest.
The case $N=2$ is a special one, because $J_2(z)=1$ - is a constant. But in this case the 2-fold
process at $\theta =\pi$ (strictly backwards) has no advantage in comparison with the direct one,
if we consider the elastic rescatterings.

The relation can be obtained from Eq. $(4.6)$
$$J_N^2(z) = J_{N-k}^2(z)J_{k+1}^2(z)-{1\over 4z^2}J_{N-k-1}^2(z)J_k^2(z) \eqno(4.9)$$
which, at $N=2m,\;k=m$ ($m$ is the integer),  leads to remarkable relation
$$ J_{2m}^2(z)=  J_m^2(z)\left(J_{m+1}^2(z) -\,{1\over 4z^2} J_{m-1}^2(z) \right) =
J_m^2(z)\left(J_{m}^2(z) -\,{1\over 2z^2} J_{m-1}^2(z) \right). \eqno(4.10) $$
Many other relations of interest can be obtained from Eq. $(4.9)$.

\section{Connection of characteristic polynomials with Chebyshev polynomials of 2-d kind.}  
The following useful relations has been found, see Eq. $(A.17)$ of \cite{koma1} (and $(6.17)$ of \cite{koma2}), which can be easily verified:
$$ \left(2z_N^\theta\right)^{N-1} J_N^2\left(z_N^\theta\right)sin{\theta\over N} = sin \theta; \qquad 
J_N^2(z_N^\theta) = {1\over \left(2z_N^\theta\right)^{N-1}} {sin \theta \over sin (\theta/ N)} \eqno(5.1) $$

It follows immediately, that zeros of $J_N(z)$ occur at $\theta = m \pi$, $m$ being any integer, so, $cos (m\pi/N)$ 
are $N-1$ roots of $J_N^2(z)$.

  Obviously, the right side of these equalities equals zero at $\theta = \pi$, but $sin (\pi/N)$ is different from zero for any integer $N\geq 2$.
Therefore, the polynomial in $cos(\pi/N)$ in the left side of $(5.1)$ should be equal to zero. 
These relations provide the link between the general case of MIP and the particular 
case of the optimal kinematics with all scattering angles equal to $\theta/N$.

The known in mathematics Chebyshev polynomials of 2-d kind \cite{cheb-engl,cheb-r} 
\footnote{According to \cite{cheb-r}, the Chebyshev polynomials of 2-d kind have been considered first by his pupils A.Korkin and E.Zolotarev 
and were named in honor of their teacher. Therefore, it is correct to name these polynomials Chebyshev-Korkin-Zolotarev, 
or CKZ-polynomials. Chebyshev polynomials are heavily used in numerical solutions (approximation theory). 
One of well known applications is in electrical filters (Chebyshev filters). }
are defined as
$$ U_n[cos\,\theta] = {sin (n+1)\theta \over sin \theta}. \eqno(5.2) $$
For any number $n$ these polynomials are defined as function of common variable $x = cos\,\theta$ which is confined in the interval
$ -1 \leq x \leq 1$. 
The recurrent relation
$$U_{n+1}(x) = 2x U_n(x) \,-\,U_{n-1}(x) \eqno(5.3) $$
can be easily checked using definition $(5.2)$.
Several examples are presented in the table.

The relation between characteristic polynomials $J^2(cos(\theta/N))$ and CKZ-polynomials $U_N(cos(\theta/N))$ takes place
$$ \left(2z_N^\theta\right)^{N-1} J_N^2(z_N^\theta) = U_{N-1} (z_N^\theta),  \eqno (5.4)$$
and recurrent relations $(5.3)$ and $(4.6)$ are identical.

Different equivalent general expressions for thr CKZ-polynomials are presented in \cite{cheb-engl}:
$$ U_n(x) = \sum_{k=0}^{[n/2]}C_{n-k}^k\, (-1)^k (2x)^{n-2k} = \sum_{k=0}^{[n/2]} C_{n+1}^{2k+1} (x^2-1)^k x^{n-2k}\quad ,  \quad n>0  \eqno(5.5) $$
where $[n/2]$ is the integer part of $n/2$, $C_n^m = n!/[m!(n-m)!]$ is the number of combinations.
The first of these formulas coincides with the expression $(4.7)$, presented in \cite{long}
up to the coefficient $[2cos(\theta/N)]^{N-1}$.
It follows from above definition $(5.2)$ that zeros (roots) of polynomials take place when $sin (n+1)\theta =0$, but  $sin \theta $ is different from
zero. So, there are $n$ roots at 
$\theta = \pi/(n+1), ...\quad \theta = n\pi / (n+1), $
and we have the equations which define the values of $cos (k\pi/n)$ at arbitrary integer $n$ and $k$.
The ortonormality conditions for the CKZ polynomials have the form
$ \int_{-1}^1 U_m(x) U_n(x) = \pi \delta_{nm}, $
which can be eazily verified using the trigonometrical definition $(5.2)$  of these polynomials.
 
\begin{center}
\begin{tabular}{|l|l|l|l|l|}                   
\hline
$N$& $J_N^2(z_N)$& $ U_{N-1} (x)$ \\
\hline
3 &$1 - 1/4x^2$  & $4x^2 -1 $                    \\
\hline
4 &$ 1-1/2x^2 $&$ 8x^3 -4 x $  \\
\hline
5 &$ 1-3/4x^2 +1/16x^4$  & $16x^4 -12 x^2 +1 $   \\
\hline
6 &$ 1-1/x^2 +3/16x^4$  & $32x^5 -32 x^3 +6x  $  \\
\hline
7 &$ 1-5/4x^2 +3/8x^4 -1 / 64 x^6$  & $64x^6 -80 x^4 + 24 x^2  -1 $  \\
\hline
8 &$ 1-3/2x^2 +5/8x^4-1 /16 x^6$  & $128x^7 -192 x^5 +80 x^3 - 8 x $ \\
\hline
9 &$ 1-7/4x^2 +15/16x^4-5 /32 x^6 +1/256 x^8 $  & $256x^8 -448 x^6 + 240 x^4 - 40 x^2 $ + 1 \\
\hline
\end{tabular}
\end{center}

{\tenrm Table. Characteristic polynomials $J_N^2 (x)$ presented in \cite{koma1,koma2} and Chebyshev 
polynomials of 2-d kind $U_{N-1}$ given in literature \cite{cheb-engl,cheb-r}.
$x=cos (\theta/N)$. The connection $ U_{N-1}(x) = (2x)^{N-1} J_N^2(x)$ can be easily verified.}\\

It has been shown in general case of the multistep process \cite{koma1,koma2} that
$$ Det \left(||a||_N\right) = {sin\,\theta/sin(\theta_1)}, \eqno(5.6) $$
where $\theta$ is the final angle of the cumulative particle.
The angle  $\theta_1$ - the polar angle of the momentum of the particle in intermediate state - is different 
from $\theta/N$ if the resonance excitation takes place. It can take place in the first interaction act, and in
any of subsequent interactions, and the optimal kinematics is changed in any of these processes.
 In any case, the resulting cross section
$$ d\sigma \sim 1/ J_N \sim \sqrt{sin(\theta_1)/sin\,\theta} \sim 1/\sqrt{\pi - \theta}, \eqno (5.7) $$
so, the backward (azimuthal) focusing effect takes place for arbitrary multistep process, but the CKZ polynomials appear for
the optimal kinematics with $\theta_k=\theta/N$, $k=1,2 ..., N-1.$ 

At singular point $\theta = \pi$ cross section of the whole MIP is proportional to $1/J_{N-1}(\pi/N)$ and is final \cite{long},
because $J_{N-1}(\pi/N)$  is different from zero, unlike $J_N(\pi/N) = 0$ \cite{kop2,long}.

\section{Conclusions}
 Physics argumentation, based on the small phase space method for 
description of the multistep processes probability in so called "kinematically forbidden regions" (cumulative 
particles production),  leads to characteristic polynomials $J_N^2[cos(\theta/N)]$, 
defining the angular dependence of cross sections in backward direction, $\theta \sim \pi$, and to the equation
$$ J_N^2[cos (\pi/N)] =0. \eqno (6.1) $$
The polynomials $J_N^2(z)$ coincide, up to some factor, with known in mathematics Chebyshev polynomials of 2-d kind, or 
Chebyshev-Korkin-Zolotqrev (CKZ) polynomials.

The nuclear glory phenomenon is a natural property
of the MIP leading to the cumulative particles production.
 The dependence 
$d\sigma \sim 1/\sqrt{\pi - \theta}$ near $\theta \sim \pi$, takes place for any
multiplicity of the process, $n\geq 3$. 
This effect, observed first at JINR and ITEP, is a clear manifestions 
of the fact that MIP make important contribution
to the cumulative particles production, although contributions
of interaction of the projectile with few-nucleon (multiquark) clusters, probably existing in nuclei, cannot be excluded.
It would be of interst and important to detect the focusing effect for different types of produced
particles, baryons and mesons (a "smoking gun" of the MIP mechanism).

\end{document}